\documentclass[11pt,preprint]{aastex}
\usepackage{psfig}

\newcommand{\citepeg}[1]{\citep[{e.g.,}][]{#1}}

\def\etal{{\sl et al.}}
\def\lsim{\hbox{ \rlap{\raise 0.425ex\hbox{$<$}}\lower 0.65ex\hbox{$\sim$} }}
\def\gsim{\hbox{ \rlap{\raise 0.425ex\hbox{$>$}}\lower 0.65ex\hbox{$\sim$} }}
\def\arcmin{\hbox{$^\prime$}}
\def\arcsec{\hbox{$^{\prime\prime}$}}

\def\f(h{\hbox{$~\!\!^{\rm h}$}}

\def\ale{\mathrel{\hbox{\rlap{\hbox{\lower4pt\hbox{$\sim$}}}\hbox{$<$}}}}
\def\age{\mathrel{\hbox{\rlap{\hbox{\lower4pt\hbox{$\sim$}}}\hbox{$>$}}}}

\slugcomment{Submitted to The Astronomical Journal}
\shortauthors{Bloom \etal~  }
\shorttitle{Optical-IR Observations of GRB 030329}
\begin{document}

\title{Optical-Infrared ANDICAM Observations of the Transient Associated with GRB 030329}

\author{J. S. Bloom\altaffilmark{1,2}, 
	P. G. van Dokkum\altaffilmark{3},
	C. D. Bailyn\altaffilmark{3}, M. M. Buxton\altaffilmark{3},
	S. R. Kulkarni\altaffilmark{4}, B. P. Schmidt\altaffilmark{5}}

\bigskip 

\affil{$^1$ Harvard Society of Fellows, 78 Mount Auburn Street, 
Cambridge, MA 02138 USA}

\affil{$^2$ Harvard-Smithsonian Center for Astrophysics, MC 20, 
60 Garden Street, Cambridge, MA 02138, USA}

\affil{$^3$ Department of Astronomy, Yale University, New Haven, CT 06520-8101, USA}

\affil{$^4$ Palomar Observatory 105--24, California Institute of Technology, Pasadena, CA 91125, USA}

\affil{$^5$ Research School of Astronomy and Astrophysics, 
Mount Stromlo Observatory, via Cotter Road, Weston, ACT 2611,
Australia}

\bigskip

\begin{abstract}

We present photometry of the transient associated with GRB\,030329
obtained with the CTIO 1.3--meter telescope and the ANDICAM
instrument, a dual optical/infrared imager with a dichroic centered at
one micron.  Without the need for light curve interpolation to produce
snapshot broadband spectra, we show that the transient spectrum
remained statistically achromatic from day 2.7 to day 5.6, during a
re-brightening episode. Associating the light in these early epochs
with the GRB afterglow, we infer a modest level of extinction due to
the host galaxy in the line--of--sight toward the GRB: $A_V$(host) =
$0.30 \pm 0.03$\,mag for $\beta = -0.5$ and $A_V$(host)\,$ < 0.4$\,mag
(3 $\sigma$) for any physically plausible value of $\beta$ (with flux
$f_\nu \propto \lambda^{-\beta}$). We conclude that the spectral slope of the
afterglow component was more than $\beta = -$0.8 between day 2.7--5.6
after the GRB, excluding the possibility that the synchrotron cooling
break passed through the optical/IR bandpass over that period. Taking
extinction into account, a decomposition of the light curve into an
afterglow and supernova component requires the presence of a supernova
similar to that of SN 1998bw, an afterglow that shows some evidence
for a second break around day 8--10, and a fifth re-brightening event
around day 15. Assuming an SN 1988bw-like evolution and a
contemporaneous GRB and SN event, the peak SN brightness was $M_V =
(-19.8 \pm 0.4) - 5\,\log_{10} h_{65}$\,mag.
\end{abstract}


\section{Introduction}

The low redshift long-duration cosmological $\gamma$-ray burst,
GRB\,030329 \citep{vcd+03} afforded an unprecedented glimpse into the
aftermath of the explosion.  The sheer brightness of the early optical
afterglow \citep{pp03,tori03} allowed for high-precision ($\sim
1-5$\%) photometric measurements to be obtained from a bevy of
meter-class telescopes \citep{uki+03,bsp+03,rs03a,pab+03,log+03}. The
optical light curve exhibited 10--30\% brightness undulations during
the first several hours \citep{uki+03} and then showed a prominent
break at day 0.48 \citep{gsb03,pfk+03}. Later, the optical transient
underwent a brightness resurgence (at about day 2) followed by an
overall decline that was punctuated by four major re-brightening
episodes
\citep{lcjf03,log03,log+03a,sysk03,gnp03}.

Beginning about 6 days after the burst, the first signs of an
underlying supernova (SN) component began to emerge in the transient,
both spectroscopically \citep{smg+03,cff+03,hsm+03,kdw+03} and
photometrically \citep{hczk03,mgs+03}; the SN was designated by the
IAU as SN 2003dh \citep{gmo+03}. The evolution of the SN spectrum in
the first two months closely parallels that of the bright type Ic SN
1998bw \citep{gvv+98,pcd+01} thus confirming the previous evidence for
(nearly) contemporaneous SN and GRB events from the same progenitor in
other bursts (see \citealt{blo03} for review).

While the nature of the SN associated with GRB\,030329 is, of course,
of intense interest, a detailed study of the afterglow holds potential
for insight into the structure of GRB explosions. It is standard
practice to model the afterglow as arising from a jetted point
explosion in a constant density environment \citepeg{rho97b,spn98} or
a wind-stratified circumburst medium \citepeg{cl99}.  However, the
assumption of an instantaneous explosion with a constant jet opening
angle is untenable both on theoretical grounds and empirical grounds.
In particular, strong variations have been seen in the early
afterglows of GRBs, e.g., GRB\,021004 \citep{fyk+03} and GRB\,030329
\citep{pfk+03,uki+03,bsp+03,rs03a,pab+03}.
 
These re-brightenings in GRB\,030329 require energy addition to the
radiating front. The energy addition can be due to slower moving
ejecta shells colliding with the earlier emitted but faster moving
ejecta
\citep{gnp03};  such a hypothesis was first proposed for the
variability in GRB 021004
\citep{fyk+03}. An alternative interpretation is that the energy
addition is due to shells with larger opening angles but with slower
moving ejecta \citep{edo03}. The placement of the behavior of the
late-time afterglow light curve of GRB\,030329 in the context of radio
modeling can help distinguish between these two hypotheses
\citep{edo03}.

An understanding of the multitude of physical processes contributing
to the complex behavior of the transient requires a careful treatment
of the high-quality data. Although analysis of the vast follow-up data
on GRB\,030329 in the literature will no doubt be of great use, we
have chosen to focus on optical-IR photometric data obtained from the
same telescope and reduced uniformly.  Whereas the burst has been
studied extensively at optical wavelengths, only one other study
presents photometric data in the near-infrared
\citep{mgs+03}. Extending the wavelength coverage by a factor of two
allows stronger constraints to be placed on the extinction, and a
better determination of the slope of the afterglow spectrum as a
function of time.

In \S \ref{sec:obs}, we present the optical and infrared data obtained
with ANDICAM between day 2 and day 23 after the GRB trigger. To
disentangle the various physical components of the transient light, in
\S \ref{sec:res} we first show that while the overall source
brightness fluctuated in the first 6 days, the spectrum of the
transient remained fixed. In \S \ref{sec:extinct}, we then fit this
early spectrum to find a constraint on the line--of--sight extinction
to the GRB and the intrinsic afterglow spectrum. Using these values,
we then decompose the afterglow and the supernova light in \S
\ref{sec:decompose}. We compare our results to those obtained in other
studies and conclude with some implications for the GRB--supernova
connection and the explosion structure of GRB\,030329.

\section{Observations, Reductions, and Calibrations}
\label{sec:obs}

All the observations reported herein were carried out with the ANDICAM
instrument mounted on the 1.3-m telescope at Cerro Tololo
Interamerican Observatory (CTIO).  The ANDICAM is a dual-channel
camera constructed by the Ohio State University instrument
group\footnotemark\footnotetext{ See {\tt
http://www.astronomy.ohio-state.edu/ANDICAM}.}.  The camera contains a
dichroic that enables two imagers to be simultaneously illuminated: a
Rockwell 1024$\times$1024 HgCdTe ``Hawaii'' array, and a Fairchild 447
2048$\times$2048 optical CCD.  This instrument was formerly mounted on
the Yale 1-m telescope at CTIO, operated by the YALO consortium
\citep{bda+99}.  ANDICAM has recently been transferred to the 1.3-m
telescope at CTIO, formerly used for the 2MASS survey, where it has
been operating since February 2003 by the SMARTS
consortium\footnotemark\footnotetext{See {\tt
http://www.astro.yale.edu/smarts}.}.  The image scale is smaller at
the 1.3-m than it was at the 1-m, and the field of view is now
2.4\arcmin $\times$2.4\arcmin\ for the IR array, and 6.3\arcmin
$\times$6.3\arcmin\ for the CCD.  The optical images are routinely
double binned to provide a pixel scale of 0.37\arcsec/pixel, while the
IR channel significantly oversamples the seeing (typically 0.8\arcsec)
with 0.14\arcsec\ pixels.

The field seen by the IR array can be repositioned slightly by
adjusting three tilt axes of an internal mirror. This allows us to
``dither'' the IR position while an optical integration is underway.
These dithered IR images can be used to generate a sky image without
actually moving the telescope.  Such sky images are slightly inferior
to those made by moving the telescope, since the field is viewed
through a different part of the filter, leading to decreased accuracy
in flat fielding.  This problem is most significant in the $K$-band,
which, at the 1.3-m, also suffers from high background levels.

Table \ref{tab:log} gives a log of the observations with ANDICAM. The
data have been grouped into 13 individual epochs. As can be seen the
total duration of the epochs was always less than 40 minutes, even
over epochs where $BVIJH$ images were acquired.  Individual optical
images were reduced using {\sc
IRAF/CCDRED}\footnotemark\footnotetext{See {\tt
http://iraf.noao.edu}. {\sc IRAF} is distributed by the National
Optical Astronomy Observatories, which are operated by the Association
of Universities for Research in Astronomy, Inc., under cooperative
agreement with the National Science Foundation.}. The bias was
removed using a mean of several bias frames and the images then
flat-fielded using the best composite twilight flats for the
appropriate filter. Twilight flats were not obtained on every night of
the transient observations but the differential gain variations over
several days should not affect the photometric uncertainty at
significantly more than the 1\% level. Images where the OT light was
contaminated with cosmic rays were excluded from the final sample.

Individual infrared images were reduced in the following way.  Flat
fields were created in $J$- and $H$-bands by subtracting pairs of
twilight flats with different sky brightnesses. After dark subtraction
each science frame was divided by the (normalized) flat field. Next, a
sky frame was created by taking the median of all normalized images
for a given epoch, and subtracted after rescaling to match the counts
in the original images. In the $H$- and $K$-bands strong ``ripples''
are present in the reduced images. These features likely result from
the ANDICAM dithering mode: as mentioned, at each mirror position the
light passes through a different part of the filter, creating
residual flat fielding errors. In the $H$-band the features could be
successfully removed by creating residual flat fields, created from
all images (at all epochs) taken at the same mirror position. In the
$K$-band the residual pattern dominated the flux of the transient even
after this correction; therefore, the $K$-band observations are not
considered further in this paper. Finally, for both the optical and IR
images, cosmic-rays were removed using a Laplacian edge-detection
algorithm \citep{vdok01}.

Usually only one or two images per epoch were obtained per optical
filter, while each epoch generated between 7--21 dithered images per
IR filter. In a given IR image, often only a few objects were detected
at better than 3 $\sigma$, requiring a complicated image alignment
procedure in order to stack the images. In a custom {\sc
Pyraf}\footnotemark\footnotetext{See {\tt
http://www.stsci.edu/resources/software\_hardware/pyraf.}} code, we
first removed any large scale gradient from a copy of all the images
using {\sc IRAF/MKSKYFIT}, masked hot and cold pixels, and then
smoothed these images with a Gaussian of width 3 pixels. The
pixel--to--pixel rms~was then determined using an iterative
sigma-clipping algorithm. This rms~was then used as an input to {\sc
DITHER/PRECOR} to mask out all regions of the sky where less than 18
out of the neighboring 5$\times$5 pixels were above three times the
rms. The effect was to create masked images where only objects appear, and
the remainder of the pixels are set to zero.
 
An initial estimate of the true offset was computed using the
positional and mirror keywords in the FITS headers. For epochs where
the seven-point dither pattern was repeated at least once, we found
that a given set of three mirror tilts (recorded in the FITS headers)
consistently mapped source positions to better than one pixel rms. From
a training set, we fit a linear least-squares transfer (3$\times$2)
matrix to map the three tilts to a nominal $x, y$ source position
taking into account the reported telescope pointing, plate
orientation, and pixel scale; we assumed no differential sky rotation
from image to image. The difference between the nominal $x, y$
position in the fiducial image and the nominal $x, y$ position in a
given image yielded an estimate of the nominal offset.

To find the precise offsets between a fiducial image and the other
images in a given epoch, we cross-correlated the masked images using
{\sc DITHER/CROSSDRIZZLE}. The highest peak in the cross correlation
within a 50 pixel width box about the nominal offset position was then
found and centroided using {\sc SHIFTFIND}; this process yielded a
list of sub-pixel offsets with uncertainties less than 0.1
pixels. Individual epochs were then stacked with {\sc DITHER/DRIZZLE}
using these offsets without re-sampling the data to a finer grid (i.e.,
{\tt drizzle.scale = 1} and {\tt drizzle.pixfrac = 1}). The offsets
between the stacked images of all IR epochs were then computed using
the same cross-correlation technique. The final epoch stacks were made
using these epoch--to--epoch offsets as input to the secondary
geometry parameters for drizzle ({\tt drizzle.dr2gpar}), repeating the
entire process described above. The result was a set of stacked,
registered images for all IR epochs.
 
Using {\sc IRAF/FITPARAM}, we calibrated our first epoch ($t = 2.67$
day) of optical images to secondary field standards reported in
\citet{hen03}. An average extinction vs.~airmass curve from nearby 
Paranal\footnotemark\footnotetext{{\tt
http://www.eso.org/instruments/isaac/imaging\_standards.html}.} was
assumed and the differential color terms found to be consistent with
zero. This implies that our cumulative filter$+$telescope response
curves reasonably approximate those of the Henden system, and thus
closely resemble the Johnson {\it BV} and Kron-Cousins {\it I} filters
(i.e., Landolt system). The photometry for all subsequent epochs were
computed differentially from the zeropoints established on the first
epoch.  As a check, primary optical Landolt standard stars were
observed on several different photometric nights and found to yield
consistent photometry to better than 5\% accuracy.

The infrared standard star SJ9144 \citep{pmk+98} was observed on the
third (photometric) epoch (April 3 UT). Using curve--of--growth
aperture photometry, we determined a zeropoint for the IR science
observations at that epoch and assume a systematic zeropoint
uncertainty of 0.05\,mag.  Aperture photometry corrected to the
``infinite aperture'' magnitudes were computed for the transient and
all detectable secondary standards in the field for all epochs. As
with the optical observations, photometry was computed differentially
from this fiducial epoch.  The resulting photometry is given in Table
\ref{tab:log}. Note that while our $J$-band photometry agrees with the 
results from Matheson et al.~to within the statistical uncertainties,
as brought to our attention by K.~Krisciunas, our $H$-band
measurements appear to be about 0.2\,mag fainter at similar epochs to
the results in Matheson et al. Unfortunately, there are no
intercomparison stars at common epochs between the two
datasets. Therefore we were unable to determine the origin of the
difference in the $H$-band.

Transient fluxes were first computed from the observed magnitudes
assuming the transformations given in \citet{fsi95} (optical) and
\citet{bb88} (IR). This revealed a small and consistent 
curvature in the broadband spectrum in the first several epochs, with
an incident spectrum ranging from $\beta \approx -1$ at the blue end
to $\beta \approx -0.6$ in the red end. Since the effective wavelength
and zeropoint of filters depend upon the incident spectrum we
recomputed these quantities by fitting a 2nd-order polynomial to the
first epoch. This fiducial spectrum was then input to {\sc
IRAF/CALCPHOT}, part of the {\sc STSDAS/SYNPHOT}
package\footnotemark\footnotetext{See {\tt
http://stsdas.stsci.edu/STSDAS.html}.} to compute the effective
wavelengths ($\lambda_{\rm eff}$) and Vega-referenced zeropoints of
the filters. The difference between fiducial filter wavelengths and
the effective filter wavelengths was small, a few tens of
Angstroms. Though the input spectrum was seen to evolve somewhat over
the course of our observations (see below), given the small changes in
$\lambda_{\rm eff}$ from the fiducial values, in the subsequent
analysis we fix $\lambda_{\rm eff}$ for all filters at all epochs. The
resulting transient fluxes are given in the last column of Table
\ref{tab:log}.

\section{Results}
\label{sec:res}

With dense high-precision sampling, the complexity of the early light
curve of GRB 030329 was revealed.  The temporal features of the early
afterglow ($t < 1$ day) have been described in detail elsewhere
\citepeg{uki+03,log03}. Later time ``bumps and wiggles'' (or
rebrightenings) are also detected in the ANDICAM light curves, shown
in Figure \ref{fig:ltcurve}.

Figure \ref{fig:spec} shows the transient broadband spectra over the
13 epochs of observations, uncorrected for the effects of dust
reddening. As evident, while the overall brightness fluctuated
(Fig.~\ref{fig:ltcurve}), there was no apparent change in the spectrum
over the first several epochs. To understand the statistical
significance of this apparent achromaticity, we take an empirical
approach by examining the changes in the 10 color indices, constructed
from the 5 observing filters (see Fig.~\ref{fig:color}). We compute
the color change from the first epoch for all subsequent epochs and
estimate the significance of change from the statistical errors on the
measurements (i.e., only the differential errors are
considered). Between epochs 1 through 4, 28 out of 30 color changes
are consistent within $\pm$2 $\sigma$ from zero change; the
significance of the $B-V$ and $B-I$ colors changes appears high on
epoch 3 (by $3$ and $4 \sigma$, respectively). From epoch 5 onward,
the distribution of color change significances fans out (as will be
shown in \S \ref{sec:decompose}, there are also secular trends in the
evolution of some colors). Specifically, the {\it Kolmogorov-Smirnov}
(KS) probability that the distribution of color change significances
from epochs 1--4 is drawn from a Gaussian of sigma unity centered
about zero is $P_{\rm KS} = 0.5$ but then plummets to below 0.05 by
day 10. We therefore conclude that the transient spectrum did not
begin to evolve significantly until after the fourth epoch of
observations.

The lack of a significant color change in the early epochs allows us
to decouple the analysis of the afterglow (dominating the first four
epochs) and the supernova (contaminating the remainder of the
epochs). In what follows, using the first four epochs only, we
determine the intrinsic afterglow spectrum and the line--of--sight
extinction concurrently. With these results, we then decompose the
afterglow component and the supernova component. In accordance with
early spectroscopic studies (e.g.,~\citealt{mgs+03}), we find that any
contribution to the total optical/IR light from an underlying
supernova was $< 5$\% in all filters during the first four epochs (see
\S \ref{sec:decompose}).

\subsection{Constraints on the Line--of--Sight Extinction}
\label{sec:extinct}

The achromaticity of the transient spectrum is consistent with the
expectation of afterglow emission in the standard synchrotron shock
model \citepeg{spn98}.  In the model, the intrinsic optical--IR
afterglow spectrum is dominated\footnotemark\footnotetext{Though the
early spectra show atmospheric absorption and star-forming emission
lines from the host \citep{smg+03,hsm+03}, these narrow features are
unlikely to dominate the wide filter light, as the equivalent width of
the lines ($\ale$ 50 \AA) is significantly smaller than the width of
the filters ($\age 1000$ \AA)} by a featureless power-law spectrum
characterized as $f_\nu \propto \lambda^{-\beta}$. There is, however, an
apparent downward curvature to the observed broadband afterglow spectrum
(Fig.~\ref{fig:spec}), which can modeled in the first four epochs as
$f_\nu \approx (-3.17 \pm 0.21) \lambda_{3}^2 + (119 \pm 4)
\lambda_{3} - (131 \pm 17)\,\mu$Jy, from $\lambda \approx 4000$ \AA\
to 18000 \AA\ ($\lambda_{3} \equiv \lambda/1000$\,\AA).

Given the expectation of an underlying power-law, the curvature is
likely due to extinction by dust along the line--of--sight to the
burst. We constrain the intrinsic extinction by normalizing the first
four epochs to a unity flux density in $I$-band and fit the function,
\begin{equation}
f_\nu(\lambda){\rm (observed)} = C \lambda^{-\beta} \times
	T[R_V{\rm (Gal)}, A_V{\rm (Gal)}, \lambda] \times 
	T[R_V{\rm (host)}, A_V{\rm (host)}, \lambda/(1 + z)]\,\mu{\rm Jy},
\end{equation}
where $R_V$ and $A_V$ are the extinction curves and equivalent
$V$-band extinction through the Galaxy (``Gal'') and the
line--of--sight through the host galaxy/progenitor system (``host''),
respectively. The redshift of the host is $z = 0.1685$
\citep{gpe+03}. A parametrized transmission function $T[R_V, A_V,
\lambda]$ is adopted from \citet{ccm89}.

The Galactic extinction toward the transient, estimated from Galactic
foreground IR emission, is $E(B-V) = 0.025$\,mag
\citep{sfd98}. Following from a recent analysis by \citet{bur03}, the
statistical error on this estimate is $\sigma_{E(B-V)} \approx
0.002$\,mag, or about 10\%. In the extinction fit we fix $R_V{\rm
(Gal)} = 3.1$ and thus fix $A_V{\rm(Gal)} = 0.083 \pm 0.017$\,mag.

By minimizing $\chi^2$, we first fit simultaneously for the values of
$C$, $R_V{\rm (host)}$, $A_V{\rm (host)}$, and $\beta$, and found the
preferred value of $R_V{\rm (host)}$ is small ($< 2$); we consider
this unlikely to be the true value of $R_V$ given the indication (from
spectroscopy and imaging) that the host is a starburst dwarf galaxy
\citepeg{mgs+03}: for such galaxies we would expect an extinction
curve $R_V \age 3$. Therefore, we examined the results for two values
of $R_V{\rm (host)} = 3.1$ and 5.1 (LMC-like). For $R_V{\rm (host)} =
3.1$ (5.1), the best fit values are $A_V{\rm (host)} = 0.94
\pm 0.24$\,mag ($1.8 \pm 1.0$\,mag), $\beta = 0.11 \pm 0.22$ 
($0.5 \pm 0.8$), $\chi^2/$dof = $0.44$ (1.34).

Note that the errors given are only derived from the diagonal elements
of the cross-correlation matrix. As Figure \ref{fig:chi} demonstrates,
there is a strong covariance between $A_V{\rm (host)}$ and $\beta$ in
the sense that an intrinsically more blue afterglow requires a higher
value of extinction. 

Though the fits given above are formally acceptable, in both cases,
the intrinsic afterglow spectrum would be unusually blue if the origin
is a relativistic synchrotron shock. In previously studied afterglows,
the observed temporal decline index $\alpha$ ($f_\nu \propto
t^\alpha$) could be used to constrain the intrinsic spectral index
through the $\alpha$--$\beta$ closure relations prescribed from
synchrotron shock physics (see \citealt{pbr+02} and references
therein). In GRB\,030329, however, the complex temporal behavior does
not yield an obvious estimation for the value of $\beta$. That said,
if we adopt the interpretation of the light curves from \citet{gnp03},
then the temporal break observed at day $0.48 \pm 0.03$ \citep{pfk+03}
can be attributed to a jet break in the standard manner, and $\beta$
then equals $-(\alpha_2 - 1)/2$. Since \citet{pfk+03} find the decay
after the break to be $\alpha_2 \approx 2$, we expect that $\beta
\approx -0.5$.

A similar value (an upper limit actually) for $\beta$ is inferred by a
consideration of the range of values of the electron energy index
spectrum $p$, where the number density of shocked electrons as a
function of energy $E$ is $n(E) \propto E^{-p}$. For the energy in the
electrons to be finite\footnotemark\footnotetext{A value of $p < 2$
would not violate the finite energy requirement as long as there
exists a high-energy cut-off in the electron energy spectrum
\citep{dc01}.}, $p$ must be greater than 2. In the simplest
homogeneous density model where $\nu_c > \nu$(opt), $\beta \le -0.5$,
consistent with value inferred above from the temporal break. The
cooling frequency $\nu_c$ corresponds to an energy at which electrons
have radiated a significant fraction of their initial energy in the
lifetime of the shock \citepeg{spn98}. We also note that the apparent
achromaticity of the afterglow over the first 4 epochs, implies that
no synchrotron break frequency (see, e.g.,
\citealt{spn98}) moved through the optical/IR bandpass from 2.7 to 5.6
days after the burst. The cooling break frequency, then, likely
remained above $\sim10^{14}$\,Hz during this time.
Imposing $\beta \le -0.5$ and $R_V$(host) $= 3.1$, we find that
$A_V$(host)\,$ \le (0.30 \pm 0.03)$\,mag. If we consider any fit
acceptable with $\chi^2$/dof $< 2$, following from Figure
\ref{fig:chi}, the extinction from the host can be zero for $p \approx
2.6$. A similar value for the extinction constraint---$A_V$(host) $\le
(0.36 \pm 0.04)$\,mag---is found for $R_V$(host) $= 5.1$. If $\beta =
-0.5$ (see Fig.~\ref{fig:best}), the combined transmission of the
Galactic and extragalactic dust is: $T_B = 0.60$, $T_V = 0.68$, $T_I =
0.77$, $T_J = 0.89$, and $T_H = 0.92$.

\citet{tmg+03} and \citet{log+03a} have noted that the X-ray to optical ratio of the afterglow remained approximately
constant from 5 hours to 60 days after the burst, with a spectral
index (if indeed connected by a single power-law) of $\beta_{\rm
opt-X}
\approx -1$. This single power-law is clearly inconsistent with our 
assertion that the optical-IR spectral slope is greater than
$-0.8$. This suggests that a (cooling) break is required between the
optical and X-ray bands above which $\beta < -1$ (see, for example,
Fig.~2 of \citealt{bfk+98}). This statement is at odds with the
conclusions of \citet{tmg+03}. If indeed, the X-ray to optical flux
ratio remained fixed during our first four epochs \citep{log+03}, this
implies that $\nu_c$ remained constant over that period. In turn, this
lends support to the hypothesis \citep{edo03} that a jet break ---
from the fast-moving ejecta responsible for optical/IR and X-ray
emission --- occurred before this time. This statement is at odds with
the conclusion of \citet{log+03a}.

\subsection{Decomposing the Afterglow and the Supernova}
\label{sec:decompose}

With an empirical examination of the color evolution in the transient,
we showed that the spectrum began to change after day $\sim$6. We can
now confirm this with a physical model for the transient, assuming a
single power-law afterglow spectrum and dust extinction.  Fixing the
values of $T_i$ found above, we fit $\beta$ as a function of time. The
result is shown in Figure \ref{fig:beta}. As seen, the first 5 epochs
are consistent with a single power-law $\beta = -0.5$, synchrotron
emission from the GRB afterglow. Then, the transient spectrum becomes
inconsistent with a single power-law as, presumably, the light from
the underlying supernova begins to contaminate.

While the spectrum on epoch 5 is formally consistent (at the 1.5
$\sigma$ level) with $\beta = -0.5$, Figure \ref{fig:spec} shows that
there was a small apparent brightening in the $V$- and $H$-bands
relative to the spectrum index between the $I$- and $J$-bands. The
$H$-band brightening could be the same ``color event'' described in
\citet{mgs+03}; the brighter $V$-band flux could be accounted for by
the onset of a non-negligible ($\sim$5--10\%) contribution from the
associated SN. In the subsequent epochs, the broadband spectrum
clearly evolves, with the light from the SN becoming an increasing
fraction of the total light at optical wavelengths (see
\citealt{mgs+03} for details of the relative contribution between the
SN and the afterglow derived from spectroscopy).

To decompose the afterglow and supernova light, we constructed a set
of predicted SN light curves in the $BVI$ bandpasses by dimming the
magnitudes of SN 1998bw to the redshift of GRB\,030329
\citep{gvv+98,pcd+01}. We calculated k--corrections derived from
1998bw spectra by using the prescription as described in
\citet{kgp96}, calculating the k--correction at each 1998bw spectral
observation epoch. A cumulative line--of--sight extinction to 1998bw
of $E(B-V)= 0.061$ was assumed \citep{wes99}. To produce a set of
synthetic SNe light curves at the (time-dilated) epoch of our
GRB\,030329 observations, these individual 1998bw epochs were fit with
a least-squares spline, with the uncertainty at any epoch gauged from
the rms scatter about the fit. The calculated k-corrections transform
the restframe (1998bw) bandpasses of $U$, $B$, $V$, $R$, $I$ to the
observed bandpasses (GRB\,030329) of $B$, $V$, $R$, $I$, $Z$,
respectively. The predicted magnitudes versus time were converted to
flux [$f_{\rm SN}(\lambda_i,t_j)$; units of $\mu$Jy]. To predict the
IR magnitudes, we fit a time-evolving blackbody spectrum to the
synthetic SN optical light curves; this temperature showed a
reasonable hard--to--soft evolution from $\log T (K^\circ)$ = 6.1 to
$\log T (K^\circ)$ = 5.8. We then converted the blackbody fluxes to
magnitudes, giving a crude estimate of the $J$- and $H$-band
magnitudes of the SN component.

For a fixed dimensionless normalization of the SN model,
$c_{\rm SN}$, we fit for the scaling of the afterglow component,
$c_{j, \rm aft}$, at each epoch $j$ where the observed flux in filter
$i$ is,
\begin{equation}
f_\nu(\lambda_{i}, t_j)[\mu{\rm Jy}] = T_i \times \left[ c_{\rm SN}\,
f_{\rm SN}(\lambda_i, t_j) + c_{j, \rm aft}(t_j)\, 
\left(\frac{\lambda_{{\rm eff}, i}}{\lambda_{{\rm eff,} I{\rm-band}}}\right)^{0.5}
\right].
\end{equation}
The first term is a scaled model of the supernova component, with
$c_{\rm SN}$ constant for all epochs. The second term is the afterglow
component, referenced to the flux at the effective wavelength of the
$I$-band filter ($\lambda_{{\rm eff,} I{\rm-band}}$; see Table 1). The
value $c_{j,
\rm aft}(t_j)$ (units of $\mu$Jy) varies as a function of time, but
the afterglow spectrum is fixed with $\beta = -0.5$. The filter
transmission values are given in \S \ref{sec:extinct}. Since pre-imaging
of the optical field of GRB\,030329 suggested a faint host galaxy
compared with the observed transient magnitudes ($R_{\rm host} >
22.5$\,mag;
\citealt{wvnl+03, bb03}) we do not include a constant host component in the fit.

Figure \ref{fig:color} shows the observed and modeled color
evolution. Without an SN component, the ``afterglow only'' model
cannot account for the secular evolution in colors, which become
manifest after about day 6. Instead, the addition of a 1998bw-like SN
--- brighter by a factor of $c_{\rm SN} \approx 1.5$ --- appears to
better represent the observed color evolution, notably from day 6 to
day 13. A value of $c_{\rm SN} = 3$ clearly over-predicts the color
changes. Had a broadband color change been due to a change in the
afterglow synchrotron spectrum itself, all color indicies should have
shown a near simultaneous change; for the passage of a cooling break,
all colors would become more red. Instead, some colors become more red
while others become more blue, consistent with a SN source spectrum
which peaks in the $V$-band.

In Figure \ref{fig:c1}, we show the fit values of $c_{j, \rm aft}$ for
three values of $c_{\rm SN}$ as well as the SN model. This analysis
demonstrates that the contribution of the SN was $\ale 5$\% at all the
epochs used to derive the extinction in \S 3.1, and hence the
determination of the values for $A_V$ and $\beta$ are not strongly
affected by the SN. In all cases, the fourth re-brightening event of
the afterglow (at $t \approx 5.2$ day; bump ``D'' in \citealt{gnp03})
is clearly seen. With a non-negligible contribution from the SN, the
decomposition reveals another steepening around 8--10 days, followed
by a possible fifth re-brightening event around 15--20 days. Our
favored valued of $c_{\rm SN} = 1.5$ implies an equipartition of the
afterglow and supernova contribution to the transient light at day
$11-13$ in the $I$-band, consistent with the spectral decomposition
from \citet{mgs+03}.

\section{Discussion and Conclusions}

Our analysis of the ANDICAM data has shown that, until day $\sim$5,
the transient light is dominated by the afterglow. During the early
epochs, the spectrum remains statistically achromatic, even during a
re-brightening event. Starting around day 6, the transient slowly
evolves in color as the supernova begins to contribute to the total
light (Fig.~\ref{fig:color} and \S \ref{sec:decompose}). In the
optical bandpasses, the early broadband spectrum is consistent with
$\beta = -0.94$, reported by \citet{smg+03} and \citet{llt+03} but is
more shallow in the IR (Fig. \ref{fig:spec}). \citet{hsm+03} report a
significantly more red spectral index of the afterglow component
($\beta = -1.2 \pm 0.05$) constrained mostly from their April 3.10
spectrum obtained on the {\it Very Large Telescope} (VLT). This is
inconsistent with our broadband spectral measurement. One explanation
for the discrepancy is that there may have been considerable
differential light loss \citep{fil82} in the VLT spectrum: over the
course of the three 600 sec exposures\footnotemark\footnotetext{See
{\tt http://archive.eso.org}.}, the parallactic angle (East of North)
changed from $\approx -170^\circ$ to $170^\circ$ yet the position
angle of the slit was fixed at 123.6$^\circ$, i.e,.~more than
40$^\circ$ from the optimal angle. Though the spectroscopic slit width
was between 1.3 to 2 times the average seeing, the combination of the
high airmass ($\sec z = 1.44$) and this chosen slit angle could easily
account for the anomalous redness of the \citet{hsm+03} spectra.

The measurement of the line--of--sight dust extinction in early GRB
afterglows is critical to determine the parameters of the shock
physics \citepeg{bdf+01}, to provide a means to deredden the flux of
any coincident supernova \citepeg{pbr+02}, and (in conjunction with an
$N_H$ measurement from X-ray spectroscopy) to constrain the extent of
dust destruction near the explosion site \citep{gw01}. Our constraint
on the line--of--sight extinction toward GRB\,030329, $A_V$(host) $\le
(0.30 \pm 0.03)$\,mag is obtained under the assumption that the
electron energy index $\beta < -0.5$. \citet{mgs+03}, using
independent broadband photometry from $U$--$H$-band on day 5.6, find a
consistent value of $A_V$(host) = $0.23 \pm 0.25$ and $\beta = -0.71
\pm 0.21$. As demonstrated in Figure \ref{fig:chi}, the coupling
between the fit values of $A_V$(host) and $\beta$ is non-negligible;
therefore, we caution that any subsequent use of these measured
quantities should take this covariance into account.  Specifically,
the lowest values of the $\chi^2$ surface are well-reproduced under
the simple constraint that $\beta = 0.92\, A_V$(host)$ - 0.75$ (for
$R_V$(host) = 3.1). Thus if $A_V$(host)$ = 0$\,mag, then $\beta =
-0.75$.

We compare the line-of-sight extinction toward GRB\,030329 with the
average extinction toward H{\sc ii} regions in the host galaxy.
Assuming case-B recombination and the \citet{ost74} interstellar
extinction curve, the H$\alpha$/H$\beta$ ratio can be converted to
$E(B-V)$ using
\begin{equation}
E(B-V)=2.21 \log \left( \frac{L_{H\alpha}}{2.76 L_{H\beta}}\right).
\end{equation}
Using Balmer line luminosities measured by \citep{hsm+03} we find
$E(B-V) = -0.01 \pm 0.25$. Taking the Galactic extinction of
$E(B-V)=0.025$ into account we find $E(B-V) < 0.23$, and
$A_V<0.76$. This is comparable to the extinction inferred by
\citet{mgs+03}. We note that the line luminosities have not been
corrected for Balmer absorption, as the host luminosity is not yet
known; however, since the equivalent widths of the Balmer lines are
large when the supernova and afterglow still dominated, this
correction will be negligible for H$\alpha$ and H$\beta$. The low
extinction in the line--of--sight toward GRB\,030329 is therefore
typical for star forming regions in this galaxy, and should therefore
not be taken as evidence for the destruction of dust by the GRB itself
\citepeg{wd00}.

As shown in Figure \ref{fig:chi}, a value of $\beta < -0.8$ is
formally excluded from the locus of acceptable fits in both cases of
$R_V$(host). This then excludes any possibility that the synchrotron
cooling break moved through the optical/IR bandpass from 2--6 days
after the GRB.  In the \S \ref{sec:decompose} we showed that the
passage of cooling break could not explain the subsequent color
changes from day 6--15, and was therefore also unlikely to have moved
through the optical/IR bandpass during that time. 

The suggested late-time afterglow features seen in Figure \ref{fig:c1}
are also an important clue for understanding the structure of GRB
explosions. \citet{edo03} suggest that the break in the radio light
curve around day 10 should have been accompanied by a second
break in the optical light curve.  Our decomposed afterglow light
curve does indeed show some evidence for such a break around day
8--10. The precise timing of the break and the slope of the post-break
afterglow light curve depends sensitively on the value of $c_{\rm
SN}$. Based on only a few later-time data points, we also suggest
tentative evidence for a fifth re-brightening event around day
15--20. If true, the mechanism for the later-time injection of energy
into the radiating front must continue at least until this time.

As shown in Figure \ref{fig:color}, a concurrent supernova about 50\%
brighter than SN\,1998bw helps to reproduce the reddening in the $B-V$
color and the bluing in the $V-I$ and $V-J$ colors. The supernova plus
afterglow model is certainly not a perfect (nor unique) fit to the
data, likely due to the inherent uncertainties in modeling the
supernova a priori.  For instance, the modeled SN over-predicts the
late-time flux at epoch 13 by $\approx 0.2-0.3$\,mag (this effect was
also noted by \citealt{hsm+03}). While the spectroscopy of the SN shows
it to be conclusively of type Ic \citep{hsm+03,cff+03,mgs+03,kdw+03},
there is a large observed diversity in color and light curves from
this class of supernovae \citepeg{imn+98,mdm+02}. SN 2003dh could
simply have risen and decayed faster than 1998bw (see also
\citealt{bkp+02}). Instead, the supernova could have occurred a few
days before the GRB (the ``supranova'' scenario;
\citealt{vs98,hsm+03}). However, as \citet{gg03} emphasize, had a GRB
occurred within a few days (or even a few months) after an SN, the
optical depth to Thompson scattering from the SN shell would be much
larger than unity, thus extinguishing the GRB. A highly asymmetric SN
explosion could produce lower optical depths for certain viewing
geometries, but the lack of strong polarization in the SN spectrum
suggests a modest level of asymmetry \citep{kdw+03}. We therefore
conclude that the SN and GRB explosions were contemporaneous to within
the free-fall collapse time of the progenitor star (see
\citealt{gg03}).

\begin{figure*}[tbp]
\centerline{\psfig{file=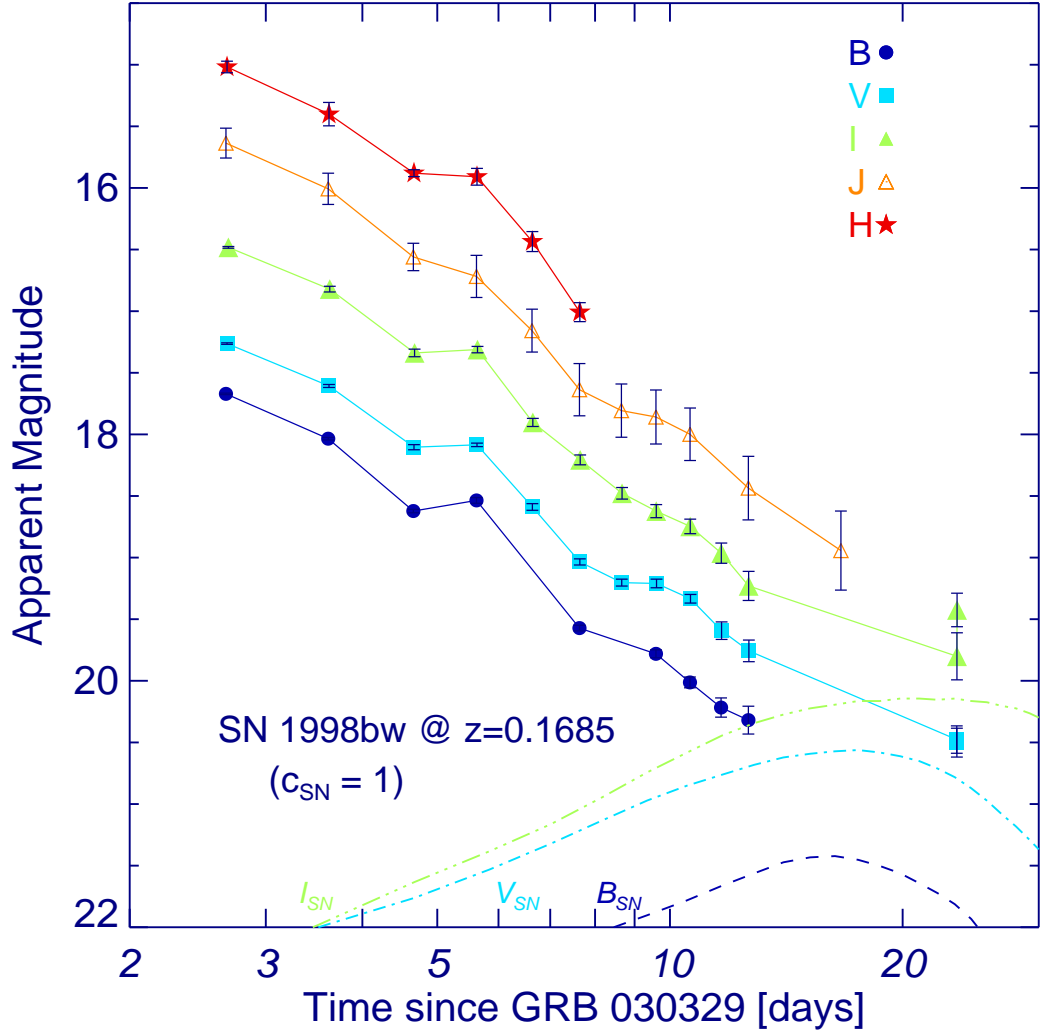,width=6.2in}}
\caption[]{
The observed light curve of the transient associated with
GRB\,030329. The complex bumps and wiggles in the early afterglow,
especially the re-brightening event at day $\sim$5.2, are
apparent. Shown are the modeled light curves in the {\it BVI} filters
(dashed, dot-dashed, dash-dotted, respectively) of SN\,1998bw at the
redshift of GRB\,030329 and dimmed by the total line--of--sight
extinction inferred for the afterglow (see text). The SN brightness
scaling ($c_{\rm SN}$) is later varied in \S \ref{sec:decompose}
to decompose the light curve into an afterglow and SN component.}
\label{fig:ltcurve}
\end{figure*}

\begin{figure*}[tbp]
\centerline{\psfig{file=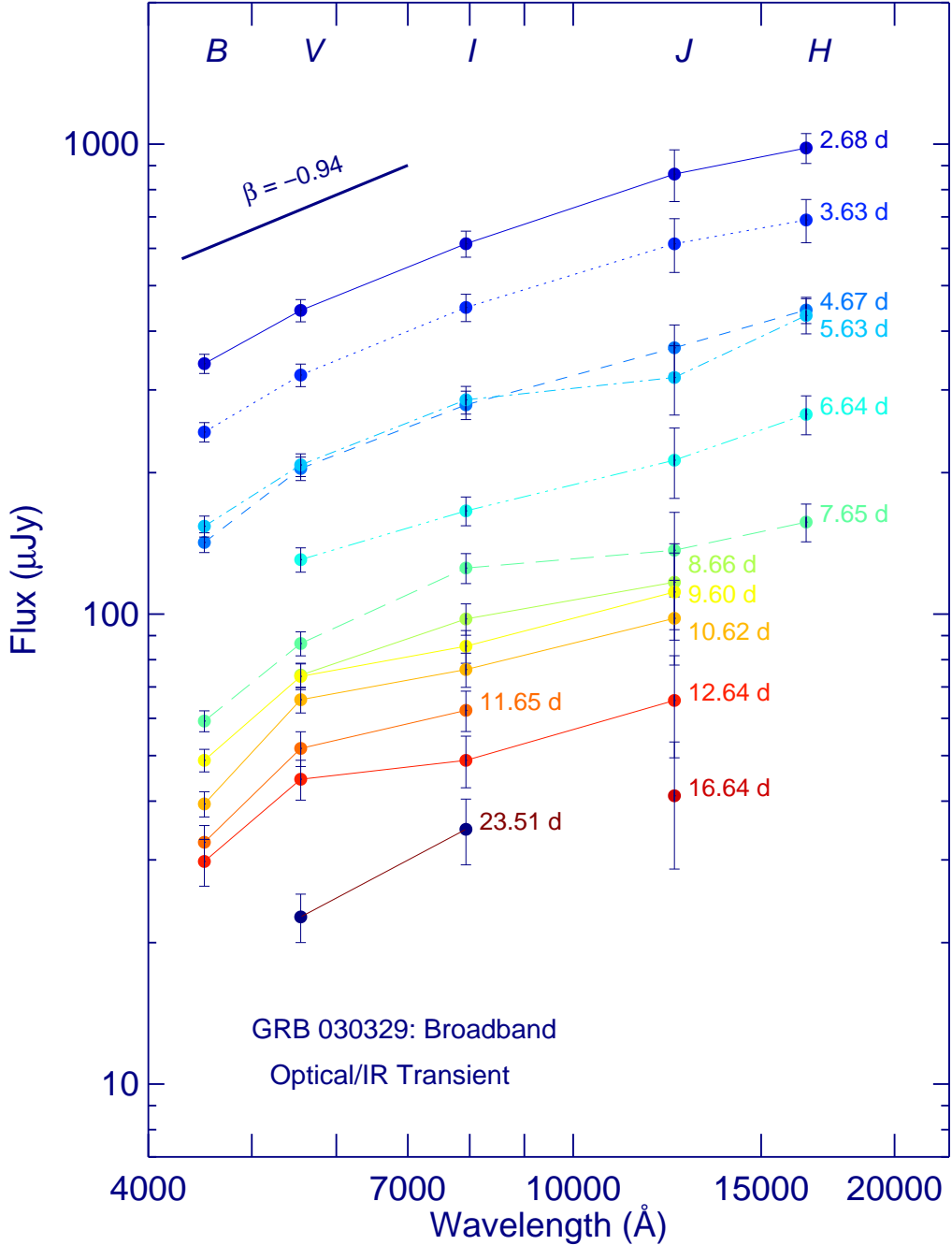,width=5.8in}}
\caption[]{ \small The broadband spectral evolution of the transient
associated with GRB\,030329. The time since the GRB is noted beside
each spectrum. The data have not been corrected for reddening. No
significant spectral evolution was detected in our first four epochs,
at least until 6 days after the GRB. For reference, the initial
optical spectral index reported by others ($\beta = -0.94$) is shown
and is consistent with our observed spectra in the bluer
filters. However, the IR data show the spectrum to flatten toward
longer wavelengths, indicative of extinction due to dust reddening. A
systematic uncertainty of 4\% in the conversion of magnitude to
flux has been included.}
\label{fig:spec}
\end{figure*}

\begin{figure*}[tbp]
\centerline{\psfig{file=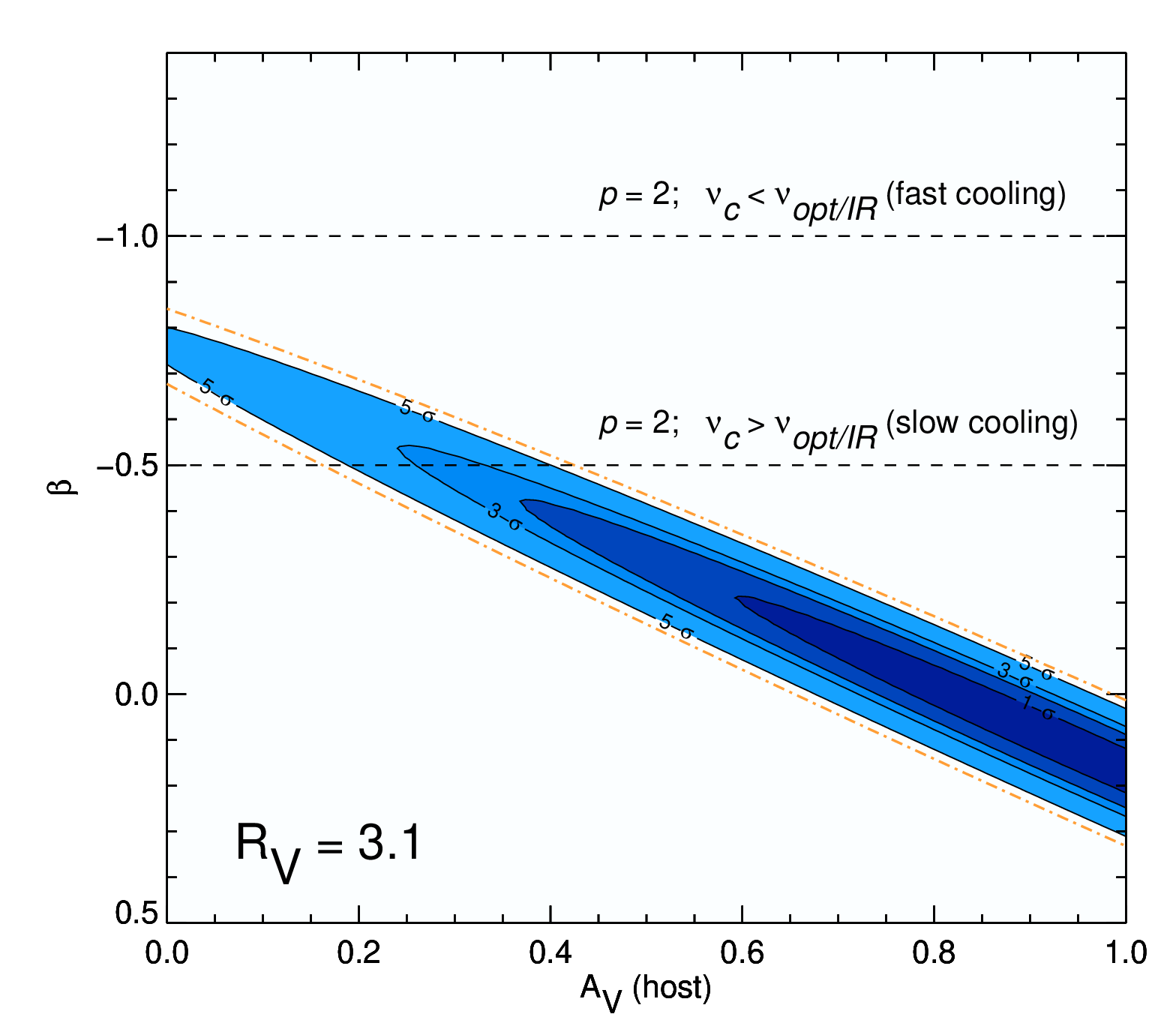,width=4.0in}}
\centerline{\psfig{file=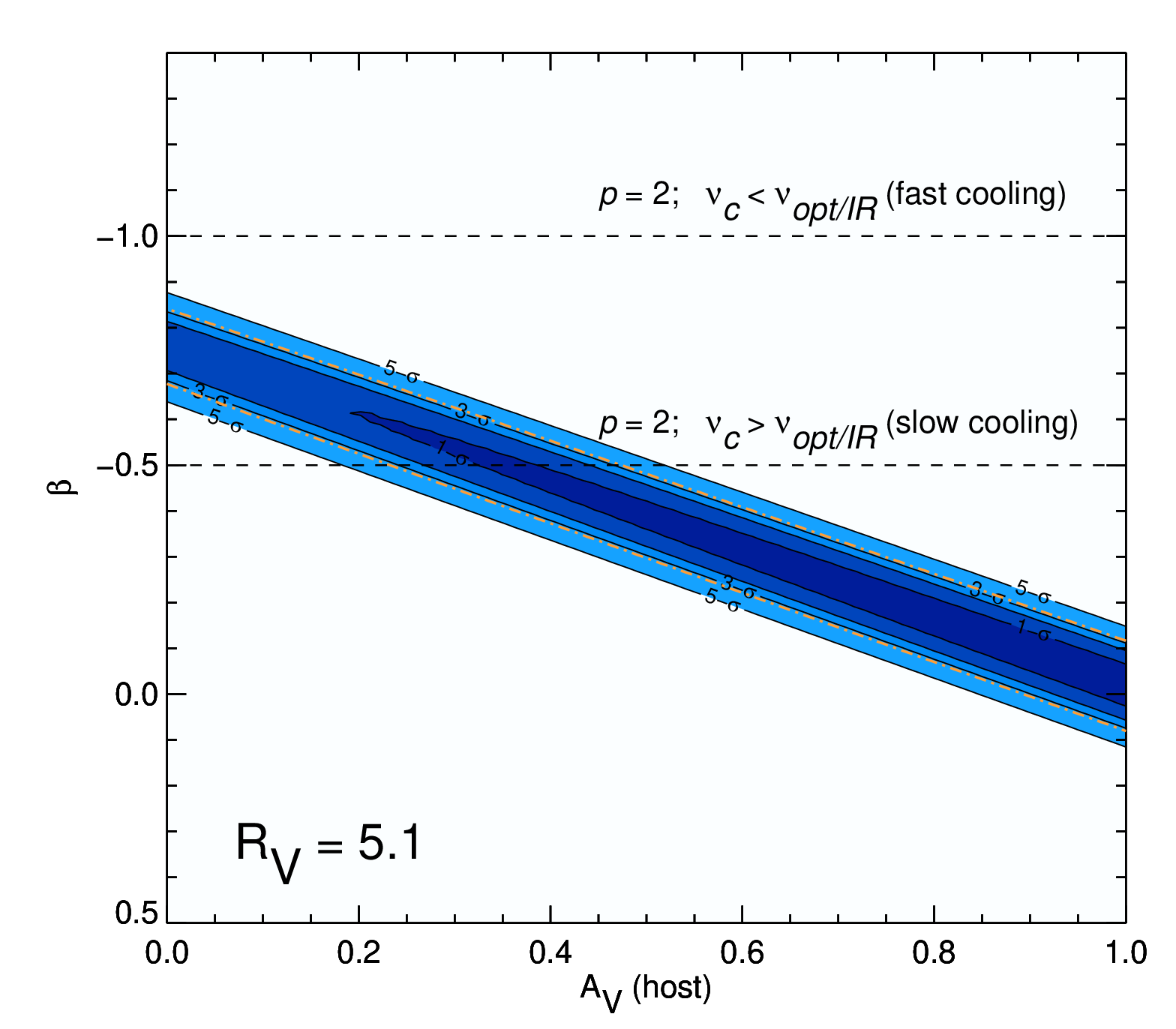,width=4.0in}}
\caption[]{
\footnotesize Coupling between the host line--of--sight restframe extinction
measurement and the intrinsic spectral power-law index $\beta$ for two
different extinction curves [top: $R_{\rm V}$(host)$ = 3.1$
(Galactic); bottom: $R_{\rm V}$(host)$ = 5.1$ (LMC)]. The first four
epochs are used in this fit, scaled to the same flux value at the
$V$-band. A Galactic extinction of $A_{\rm V}$(Gal)$ = 0.083$\,mag,
$R_{\rm V}$(Gal)$ = 3.1$ is assumed.  Lower values of host extinction
spectra result in lower (more red) values of $\beta$. The data prefer
a blue afterglow with high host extinction, though physical
constraints require a more red afterglow: the horizontal dashed lines
show the predicted $\beta$ for $p = 2$ in the fast and slow cooling
regimes. For $p > 2$, in either regime, $\beta$ must be more
negative. Solid contours show the 1-, 2-, 3-, and 5-$\sigma$ reflect
the 2-parameter demarcations from the $\chi^2$ surface. The
dash-dotted line shows an acceptable region of the data, where
$\chi^2$/dof $< 2$. Therefore, $\beta$ must be greater than $-0.8$ in
the first four epochs, ruling out the hypothesis that the cooling
break, $\nu_c$, passed through the optical/IR bandpass between days 2
and 6.}
\label{fig:chi}
\end{figure*}

\begin{figure*}[tbp]
\centerline{\psfig{file=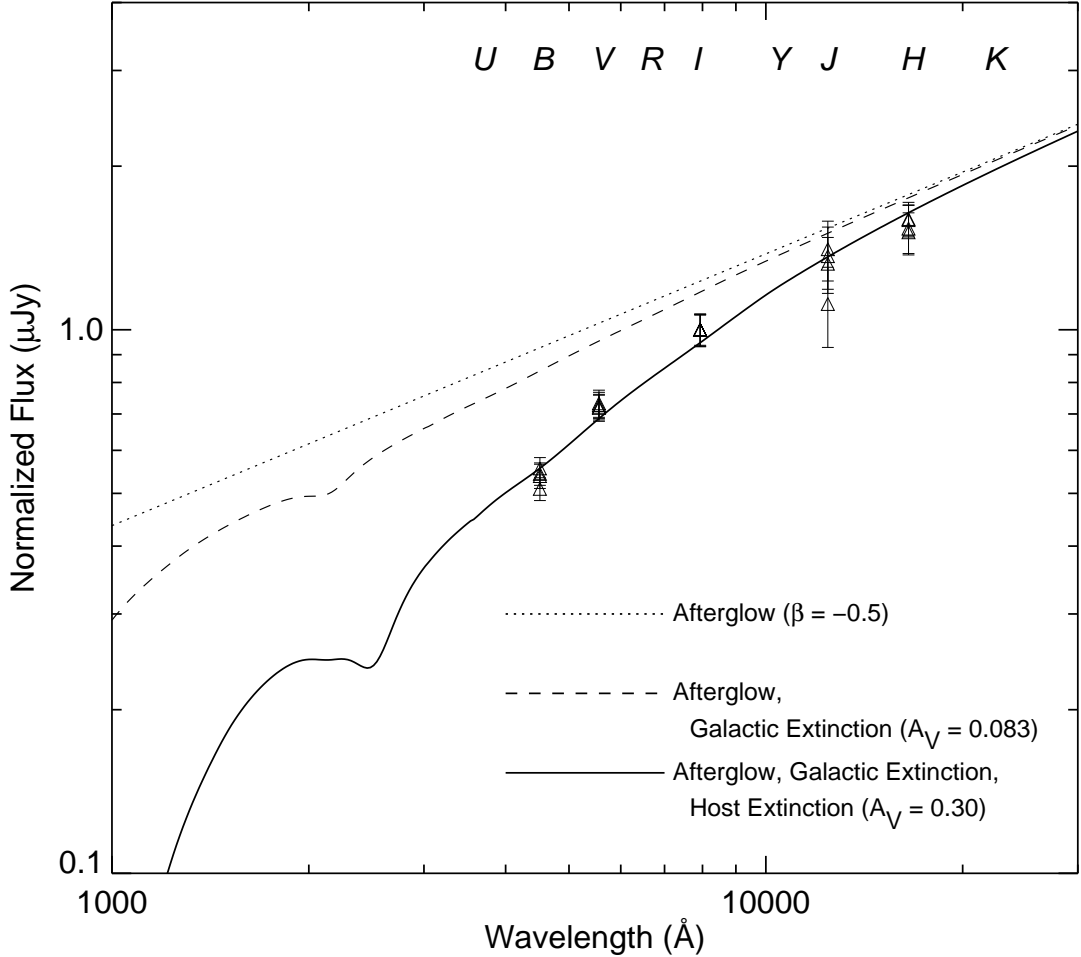,width=6.2in}}
\caption[]{
Best-fit afterglow spectrum plus extinction assuming $\beta = -0.5$
($p=2$, slow cooling). As in Figure \ref{fig:chi}, the Galactic
extinction is fixed. The dotted line shows the unreddened afterglow
spectrum. The dashed line shows the afterglow spectrum reddened only
with Galactic extinction. The solid line shows the afterglow after
reddening from the Galaxy and a host line--of--sight extinction of
$A_{\rm V} = 0.30 \pm 0.03$ mag (assuming $R_{\rm V}$(host)\,$ =
3.1$). This fit is acceptable, with $\chi^2$/dof $=0.94$.}
\label{fig:best}
\end{figure*}

\begin{figure*}[tbp]
\centerline{\psfig{file=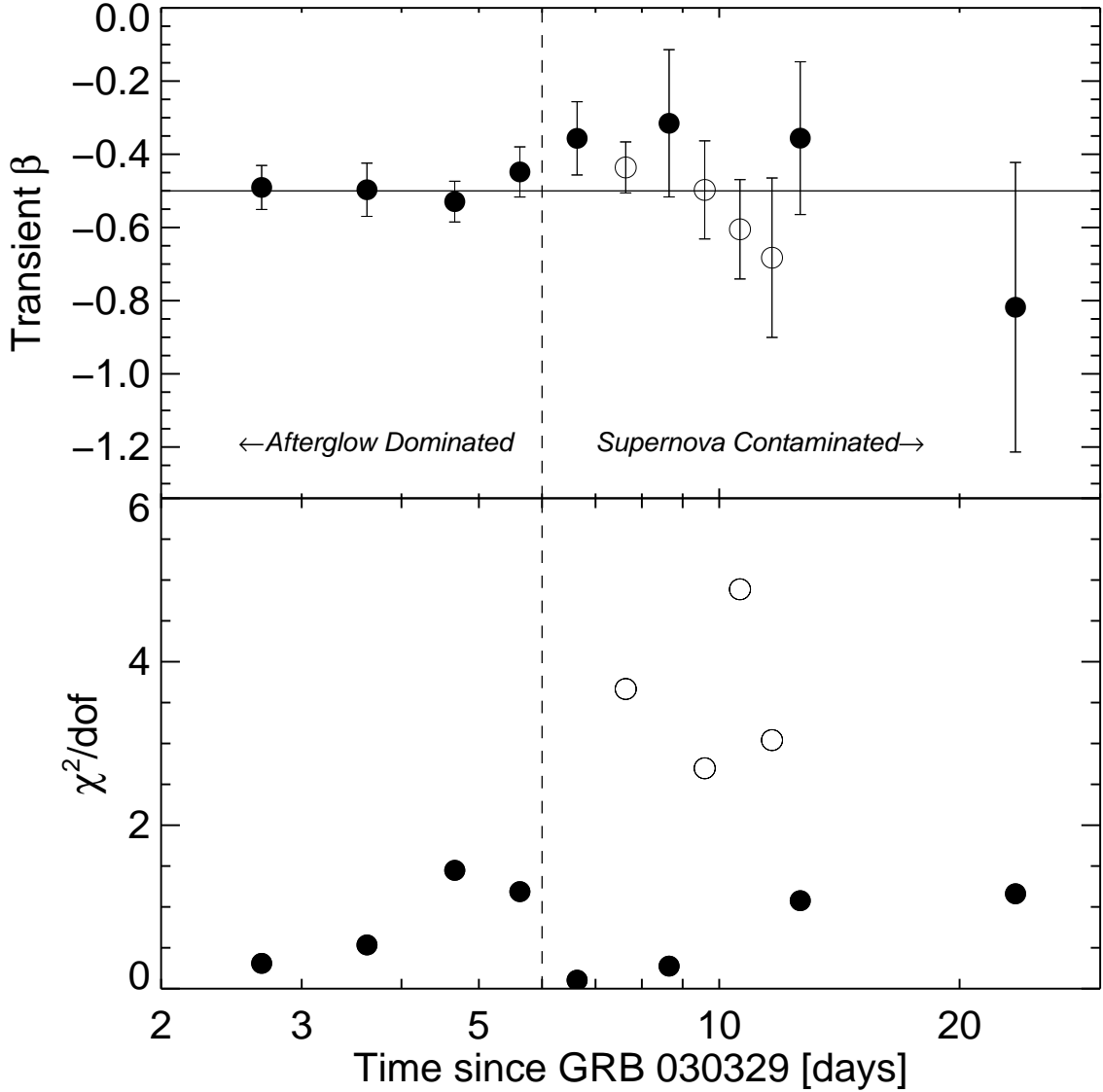,width=6.2in}}
\caption[]{ Power-law fit to the evolving intrinsic transient spectrum
as a function of time, fixing the extinction measured in
Fig.~\ref{fig:best}. The top panel gives the value for $\beta$ and the
associated 1 $\sigma$ error assuming an acceptable fit. The bottom
panel gives the $\chi^2$/dof of the fit. Measurements at epochs marked
with solid circles are acceptable fits ($\chi^2$/dof\,$< 2$), implying
that the transient can be adequately modeled as a single
power-law. Open circles show unacceptable fits to a single power-law.}
\label{fig:beta}
\end{figure*}

\begin{figure*}[tbp]
\centerline{\psfig{file=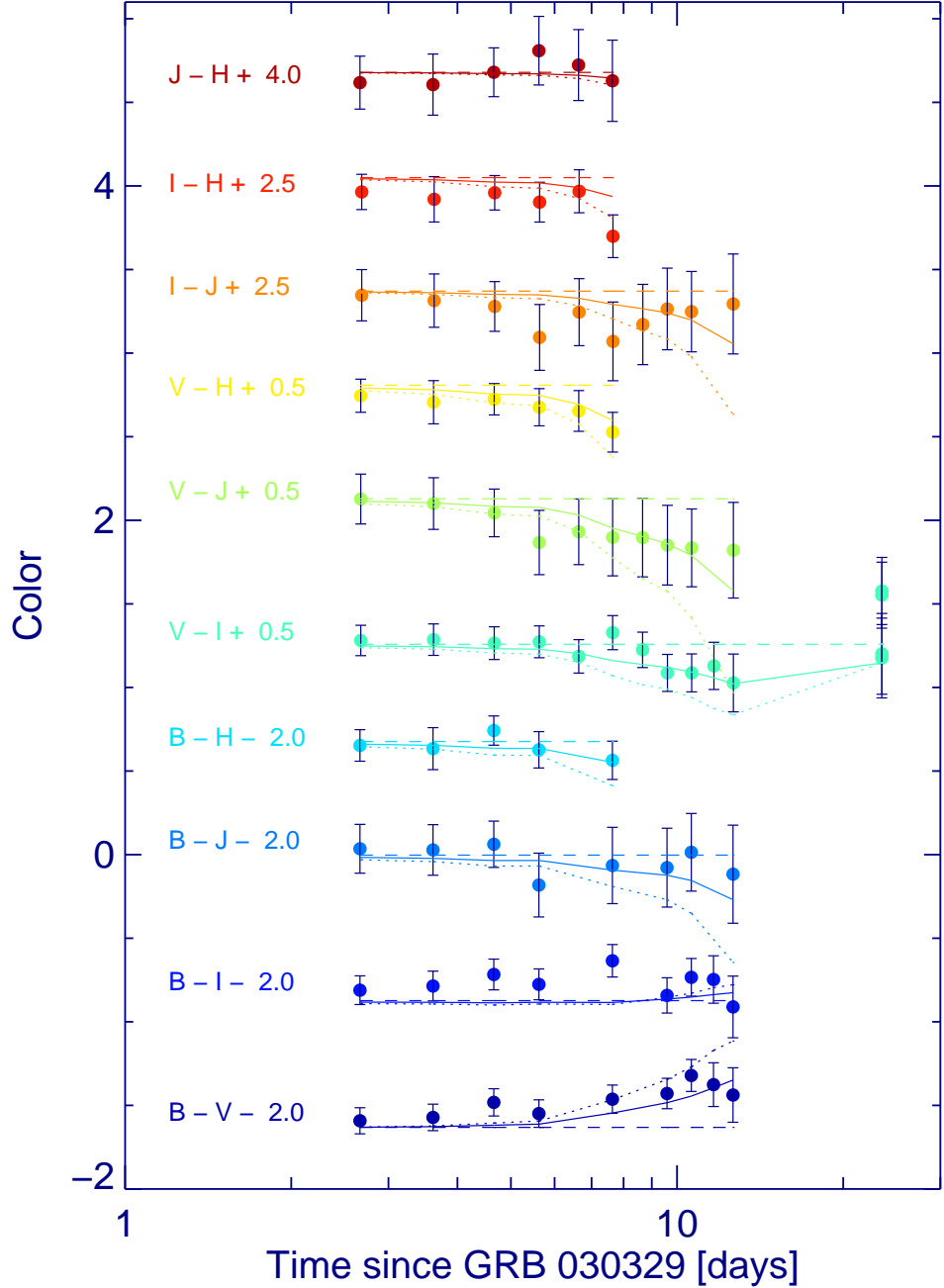,width=5.7in}}
\caption[]{
\footnotesize
Color evolution of the observed transient compared with a model for
the afterglow and the supernova component. The observed data,
uncorrected for dust extinction, is shown with 1-$\sigma$ error
bars. The model, using a mix of an afterglow and a 1998bw-like
supernova, was constructed following \S \ref{sec:decompose}. The
dashed horizontal line derives from the best fit result to $c_{j, \rm
aft}$ assuming no contribution from a supernova (i.e., $c_{\rm SN} =
0$). The solid (dotted) lines show the color results for $c_{\rm SN} =
1.5$ (3.0). The $c_{\rm SN} = 1.5$ helps reproduce the apparent bluing
trend in $V - I$, $V - J$, and $V - H$ and the reddening trend in $B -
V$. Such color trends are a natural consequence of an SN source
spectrum which peaks around 5000--6000 \AA.}
\label{fig:color}
\end{figure*}

\begin{figure*}[tbp]
\centerline{\psfig{file=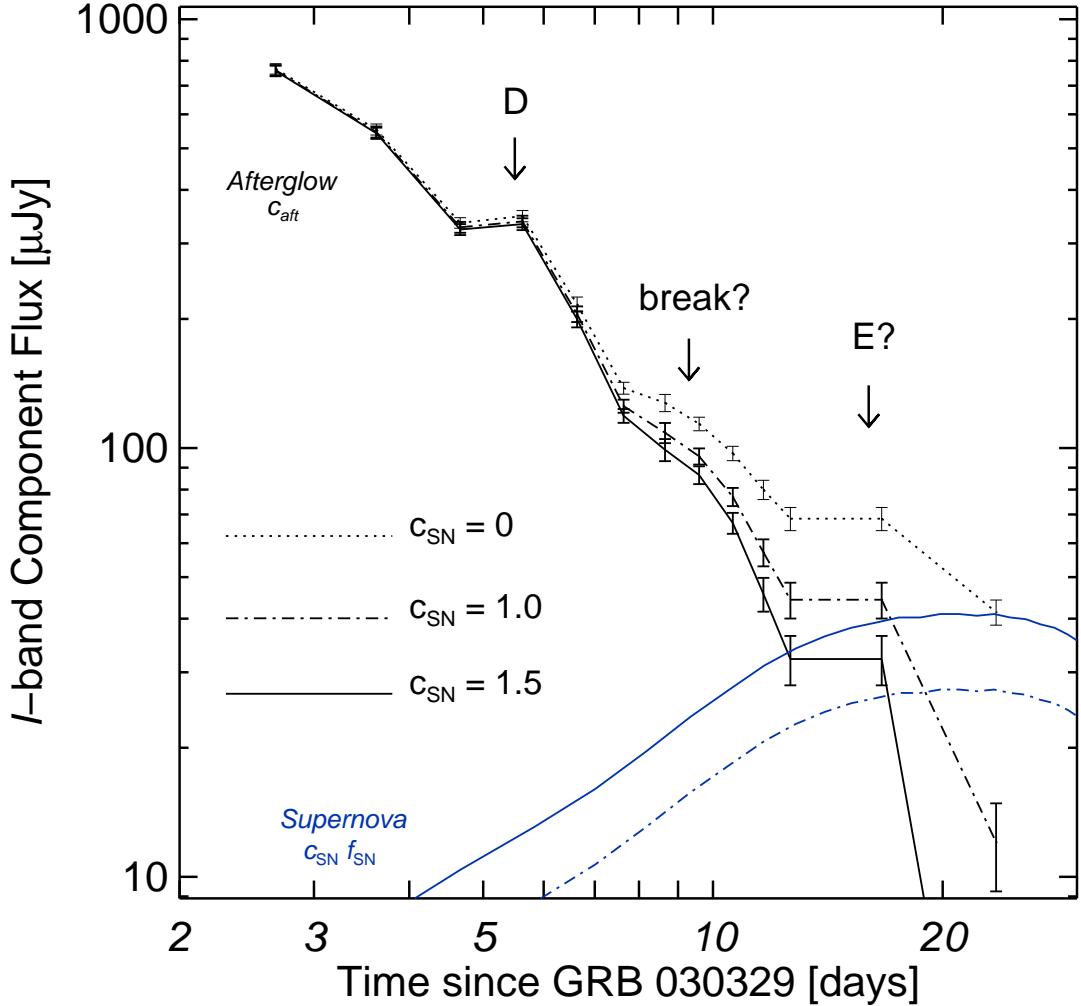,width=6.2in}}
\caption[]{ Evolution of the afterglow and supernova brightness at the
effective wavelength of the $I$-band filter for different levels of
contribution from an underlying supernova. The temporal dependence on
the scaling parameter $c_{j, \rm aft}$ is shown as a dotted curve
assuming no contribution from a supernovae ($c_{\rm SN} = 0$); in
Figure \ref{fig:color} we show that a model with $c_{\rm SN} = 0$ does
not reproduce the color evolution adequately. Instead, a reasonable
agreement in both color and flux is obtained by setting $c_{\rm SN} =
1.5$. The sum of both components for a given $c_{\rm SN}$ should equal
that of the dotted curve within the errors. The fourth re-brightening
event (``D'' from \citealt{gnp03}) is labeled with an arrow at day
5.2. There is some evidence for a break in the light curve between day
$\sim$8--10 and another re-brightening event in the light curve around
day 15--20, labeled as ``E''. The fluxes have been corrected for the
effects of dust extinction.}
\label{fig:c1}
\end{figure*}

Since we did not observe during the peak of the SN (see Figure
\ref{fig:ltcurve}), we do not have a direct measurement of the peak SN
brightness. However, assuming a 1998bw-like evolution of the supernova
component, scaled to be 1.5 times brighter (0.44\,mag), the peak absolute
magnitude of 2003dh is $M_V = -19.8 - 5\,\log_{10} h_{65}$\,mag; we
estimate an uncertainty in this value due to $c_{\rm SN}$ of 0.3\,mag.
There is an additional $\sim$10\% systematic uncertainty in this
quantity owing to the unknown peculiar velocity of the host of
SN\,1998bw. 

We end by highlighting one of the unique features of ANDICAM, namely
the ability to simultaneously image afterglows in both the optical and
infrared bands. Aside from allowing for efficient characterization
of the broadband spectrum to measure line--of--sight extinction, any
short timescale ($\sim$minutes) color changes in the afterglow could
be detected unambiguously. If such observations were to be conducted
in the first few hours after a GRB, strong constraints on the passage
of the synchrotron peak frequency could be obtained, leading to
precision measurements of the peak Lorentz factor of the shell. In
addition, the constraints on short term color variations would lead to
constraints on the patchiness of clouds near the progenitor
\citep{kp00a}. Indeed, with such instruments in the {\it Swift} era we 
look forward to a level of insight even beyond that gleaned from
observations of the remarkable transient of GRB\,030329.


\acknowledgments

This paper reports data taken through Yale University's share of the
SMARTS consortium. We extend our gratitude to K.~Krisciunas for
helpful discussions and for providing us with details of the IR
analysis in the Matheson paper. We thank the ANDICAM operator,
D.~Gonzalez, for his dedication to observing this source.  We also
thank S.~Barthelmy and the GCN team. The feedback and help with
photometry from A.~Gal-Yam has been appreciated. C.~D.~B.~and
M.~M.~B.~are supported by NSF grant AST-0098421. J.~S.~B.~is supported
by a Junior Fellowship to the Harvard Society of Fellows and by a
generous research grant from the Harvard-Smithsonian Center for
Astrophysics.

\newpage

\newpage

\begin{deluxetable}{lcccccl}
\tabletypesize{\footnotesize}
\tablecaption{Photometric Observations of GRB\,030329\label{tab:log}}
\tablecolumns{7}
\label{tab:param}
\tablehead{
\colhead{Date\tablenotemark{a}} & \colhead{$\Delta t$\tablenotemark{b}} & \colhead{Filter} &
\colhead{Exp.~Time} & \colhead{Airmass} &
\colhead{Mag.\tablenotemark{c}} & \colhead{Flux\tablenotemark{d}} \\
\colhead{UT} & \colhead{day} & \colhead{} &
\colhead{sec} & \colhead{$\sec z$} & \colhead{} &
\colhead{$\mu$Jy} \\
\colhead{(1)} & \colhead{(2)} & \colhead{(3)} &
\colhead{(4)} & \colhead{(5)} & \colhead{(6)} &
\colhead{(7)}}
\startdata
\cutinhead{Epoch        1, $\delta t= 38.5$ min}
Apr        1 03:33:29 &  2.664 & {\rm B} &  600.00 & 1.64 & $ 17.67 \pm  0.01$ & $  341.1 \pm     9.9$ \\
Apr        1 03:45:52 &  2.673 & {\rm V} &  600.00 & 1.67 & $ 17.26 \pm  0.01$ & $  442.8 \pm    17.7$ \\
Apr        1 04:01:30 &  2.684 & {\rm I} &  600.00 & 1.71 & $ 16.48 \pm  0.01$ & $  613.8 \pm    31.9$ \\
Apr        1 03:33:27 &  2.664 & {\rm J} &  560.19 & 1.64 & $ 15.64 \pm  0.12$ & $  863.5 \pm   104.1$ \\
Apr        1 03:45:50 &  2.673 & {\rm H} &  560.18 & 1.67 & $ 15.02 \pm  0.05$ & $  981.2 \pm    62.0$ \\
\cutinhead{Epoch        2, $\delta t= 35.1$ min}
Apr        2 02:19:40 &  3.613 & {\rm B} &  600.00 & 1.62 & $ 18.03 \pm  0.01$ & $  243.9 \pm     7.3$ \\
Apr        2 02:31:56 &  3.621 & {\rm V} &  600.00 & 1.61 & $ 17.61 \pm  0.01$ & $  322.6 \pm    13.3$ \\
Apr        2 02:44:17 &  3.630 & {\rm I} &  600.00 & 1.61 & $ 16.82 \pm  0.02$ & $  449.2 \pm    25.0$ \\
Apr        2 02:19:38 &  3.613 & {\rm J} &  560.19 & 1.62 & $ 16.01 \pm  0.13$ & $  613.6 \pm    77.1$ \\
Apr        2 02:31:53 &  3.621 & {\rm H} &  560.18 & 1.61 & $ 15.40 \pm  0.09$ & $  689.5 \pm    68.2$ \\
\cutinhead{Epoch        3, $\delta t= 34.9$ min}
Apr        3 03:21:50 &  4.656 & {\rm B} &  600.00 & 1.63 & $ 18.62 \pm  0.01$ & $  142.1 \pm     4.5$ \\
Apr        3 03:34:01 &  4.664 & {\rm V} &  600.00 & 1.66 & $ 18.10 \pm  0.02$ & $  204.1 \pm     9.0$ \\
Apr        3 03:46:14 &  4.673 & {\rm I} &  600.00 & 1.69 & $ 17.34 \pm  0.03$ & $  278.7 \pm    16.4$ \\
Apr        3 03:21:48 &  4.656 & {\rm J} &  480.15 & 1.63 & $ 16.56 \pm  0.11$ & $  368.7 \pm    41.3$ \\
Apr        3 03:33:58 &  4.664 & {\rm H} &  560.20 & 1.66 & $ 15.88 \pm  0.03$ & $  443.6 \pm    23.4$ \\
\cutinhead{Epoch        4, $\delta t= 34.9$ min}
Apr        4 02:29:21 &  5.620 & {\rm B} &  600.00 & 1.61 & $ 18.54 \pm  0.02$ & $  153.8 \pm     5.5$ \\
Apr        4 02:41:30 &  5.628 & {\rm V} &  600.00 & 1.61 & $ 18.08 \pm  0.01$ & $  207.7 \pm     8.7$ \\
Apr        4 02:53:44 &  5.636 & {\rm I} &  600.00 & 1.61 & $ 17.31 \pm  0.03$ & $  285.8 \pm    16.3$ \\
Apr        4 02:29:18 &  5.619 & {\rm J} &  560.18 & 1.61 & $ 16.72 \pm  0.17$ & $  318.8 \pm    52.3$ \\
Apr        4 02:41:28 &  5.628 & {\rm H} &  560.18 & 1.61 & $ 15.91 \pm  0.07$ & $  431.9 \pm    33.6$ \\
\cutinhead{Epoch        5, $\delta t= 33.2$ min}
Apr        5 02:55:21 &  6.638 & {\rm V} &  600.00 & 1.61 & $ 18.59 \pm  0.03$ & $  130.6 \pm     6.1$ \\
Apr        5 03:07:36 &  6.646 & {\rm I} &  600.00 & 1.63 & $ 17.90 \pm  0.03$ & $  165.8 \pm     9.9$ \\
Apr        5 02:44:50 &  6.630 & {\rm J} &  480.17 & 1.61 & $ 17.16 \pm  0.17$ & $  212.5 \pm    35.4$ \\
Apr        5 02:55:19 &  6.638 & {\rm H} &  560.19 & 1.61 & $ 16.43 \pm  0.08$ & $  266.0 \pm    23.3$ \\
\cutinhead{Epoch        6, $\delta t= 34.6$ min}
Apr        6 02:54:10 &  7.637 & {\rm B} &  600.00 & 1.62 & $ 19.57 \pm  0.02$ & $   59.2 \pm     2.1$ \\
Apr        6 03:06:13 &  7.645 & {\rm V} &  600.00 & 1.63 & $ 19.03 \pm  0.03$ & $   86.6 \pm     4.0$ \\
Apr        6 03:18:20 &  7.654 & {\rm I} &  600.00 & 1.65 & $ 18.21 \pm  0.04$ & $  125.3 \pm     7.9$ \\
Apr        6 02:54:08 &  7.637 & {\rm J} &  560.17 & 1.62 & $ 17.64 \pm  0.21$ & $  136.7 \pm    27.4$ \\
Apr        6 03:06:11 &  7.645 & {\rm H} &  560.21 & 1.63 & $ 17.01 \pm  0.08$ & $  156.9 \pm    13.3$ \\
\cutinhead{Epoch        7, $\delta t= 34.7$ min}
Apr        7 03:30:34 &  8.662 & {\rm V} &  600.00 & 1.69 & $ 19.20 \pm  0.03$ & $   74.2 \pm     3.5$ \\
Apr        7 03:42:45 &  8.670 & {\rm I} &  600.00 & 1.72 & $ 18.48 \pm  0.05$ & $   97.6 \pm     6.6$ \\
Apr        7 03:18:28 &  8.654 & {\rm J} &  560.18 & 1.66 & $ 17.81 \pm  0.22$ & $  116.9 \pm    23.9$ \\
\cutinhead{Epoch        8, $\delta t= 34.6$ min}
Apr        8 01:48:29 &  9.591 & {\rm B} &  600.00 & 1.63 & $ 19.78 \pm  0.03$ & $   48.9 \pm     2.0$ \\
Apr        8 02:12:35 &  9.608 & {\rm V} &  600.00 & 1.61 & $ 19.21 \pm  0.04$ & $   73.7 \pm     3.8$ \\
Apr        8 02:00:33 &  9.600 & {\rm I} &  600.00 & 1.62 & $ 18.62 \pm  0.05$ & $   85.4 \pm     6.1$ \\
Apr        8 01:48:26 &  9.591 & {\rm J} & 1800.18 & 1.63 & $ 17.86 \pm  0.22$ & $  111.4 \pm    23.1$ \\
\cutinhead{Epoch        9, $\delta t= 34.5$ min}
Apr        9 02:21:09 & 10.614 & {\rm B} &  600.00 & 1.61 & $ 20.01 \pm  0.04$ & $   39.5 \pm     2.0$ \\
Apr        9 02:45:13 & 10.631 & {\rm V} &  600.00 & 1.62 & $ 19.33 \pm  0.04$ & $   65.7 \pm     3.4$ \\
Apr        9 02:33:09 & 10.622 & {\rm I} &  600.00 & 1.61 & $ 18.75 \pm  0.06$ & $   76.2 \pm     5.7$ \\
Apr        9 02:25:49 & 10.617 & {\rm J} & 1680.18 & 1.61 & $ 18.00 \pm  0.21$ & $   98.0 \pm    19.7$ \\
\cutinhead{Epoch       10, $\delta t= 36.4$ min}
Apr       10 03:05:21 & 11.645 & {\rm B} &  600.00 & 1.65 & $ 20.22 \pm  0.08$ & $   32.7 \pm     2.5$ \\
Apr       10 03:31:19 & 11.663 & {\rm V} &  600.00 & 1.72 & $ 19.59 \pm  0.07$ & $   51.8 \pm     4.0$ \\
Apr       10 03:17:38 & 11.653 & {\rm I} &  600.00 & 1.68 & $ 18.96 \pm  0.08$ & $   62.4 \pm     5.7$ \\
\cutinhead{Epoch       11, $\delta t= 36.0$ min}
Apr       11 02:47:16 & 12.632 & {\rm B} &  600.00 & 1.63 & $ 20.32 \pm  0.11$ & $   29.8 \pm     3.2$ \\
Apr       11 03:12:50 & 12.650 & {\rm V} &  600.00 & 1.68 & $ 19.76 \pm  0.09$ & $   44.5 \pm     4.0$ \\
Apr       11 03:00:10 & 12.641 & {\rm I} &  600.00 & 1.65 & $ 19.23 \pm  0.12$ & $   48.8 \pm     5.9$ \\
Apr       11 02:47:14 & 12.632 & {\rm J} & 1800.19 & 1.63 & $ 18.44 \pm  0.26$ & $   65.5 \pm    15.9$ \\
\cutinhead{Epoch       12, $\delta t= 33.4$ min}
Apr       15 02:58:11 & 16.640 & {\rm J} & 1920.15 & 1.68 & $ 18.94 \pm  0.32$ & $   41.1 \pm    12.3$ \\
\cutinhead{Epoch       13, $\delta t= 36.2$ min}
Apr       21 23:36:46 & 23.500 & {\rm V} &  450.00 & 1.90 & $ 20.48 \pm  0.11$ & $   22.9 \pm     2.5$ \\
Apr       21 23:46:05 & 23.506 & {\rm V} &  450.00 & 1.85 & $ 20.50 \pm  0.12$ & $   22.4 \pm     2.6$ \\
Apr       21 23:55:43 & 23.513 & {\rm I} &  450.00 & 1.80 & $ 19.80 \pm  0.19$ & $   28.9 \pm     5.3$ \\
Apr       22 00:05:07 & 23.519 & {\rm I} &  450.00 & 1.76 & $ 19.42 \pm  0.14$ & $   40.8 \pm     5.5$ \\
\enddata 

\tablecomments{\footnotesize These photometric observations have been grouped into 
thirteen epochs. The total duration of the epoch --- from the time of
the beginning of the first exposure until the ending time of the last
exposure --- is shown as $\delta t$.}

\tablenotetext{a}{\footnotesize UT start time of the exposure. All observations were conducted in the year 2003.}

\tablenotetext{b}{\footnotesize Time since the GRB trigger.}

\tablenotetext{c}{\footnotesize Observed magnitude in the Landolt+Persson filter system, 
uncorrected for Galactic and host reddening. The errors given are
statistical only. The systematic uncertainty in the zeropoint
calibrations, to be added in quadrature with the statistical errors,
are $B$=0.03\,mag, $V$=0.04 mag, $I$=0.06\,mag, $J$=0.05\,mag, and
$H$=0.05\,mag.}

\tablenotetext{d}{\footnotesize Equivalent flux measurement at the effective 
wavelength of the respective filter ($\lambda_{\rm eff}$($B$) =
4513.5 \AA, ($V$) = 5556.3 \AA, ($I$) = 7935.0 \AA, ($J$) = 12440 \AA,
($H$) = 16528 \AA). As described in the text, the values for
$\lambda_{\rm eff}$ and zeropoints were determined self-consistently
for the observed input spectrum. No correction due to reddening from
dust has been applied. The errors reflect both the statistical and
systematic magnitude zeropoint uncertainty in the flux measurement.  No
uncertainty in the filter flux zeropoint has been included.}

\end{deluxetable}

\end{document}